%% file: main.tex
\documentclass[sigconf]{acmart}
\AtBeginDocument{
  \providecommand\BibTeX{{
    \normalfont B\kern-0.5em{\scshape i\kern-0.25em b}\kern-0.8em\TeX}}}

\input{packages}
\input{commands}

\setcopyright{acmcopyright}
\copyrightyear{2021}
\acmYear{2021}
\acmDOI{}
\acmConference[MMSys'21]{12th ACM Multimedia
Systems Conference}{Sept. 28-Oct. 1, 2021}{Istanbul, Turkey}
\acmBooktitle{12th ACM Multimedia Systems Conference (MMSys'21), Sept. 28-Oct. 1, 2021,
2021, Istanbul, Turkey}
\acmPrice{15.00}
\acmISBN{}

\begin{document}

\title{MMSys'21 Grand Challenge on Detecting Cheapfakes}
\input{authors}
\input{sections/0-abstract}
\input{figures/teaser-figure}

\maketitle

\input{sections/1-introduction}
\input{sections/2-dataset}
\input{sections/3-task}
\input{sections/4-evaluation}
\input{sections/5-administrative}
\input{sections/6-conclusion}

\bibliographystyle{ACM-Reference-Format}
\bibliography{references.bib}

\end{document}

%% file: packages.tex
\usepackage[nolist,nohyperlinks]{acronym}
\acrodef{E1}{Effectiveness Score}
\acrodef{E2}{Efficiency Score}
\acrodef{FN}{False Negatives}
\acrodef{FP}{False Positives}
\acrodef{JSON}{JavaScript Object Notation}
\acrodef{OOC}{out-of-context}
\acrodef{MCC}{Matthews Correlation Coefficient}
\acrodef{NOOC}{not-out-of-context}
\acrodef{PDF}{Portable Document Format}
\acrodef{TN}{True Negatives}
\acrodef{TP}{True Positives}

\usepackage{listings}
\usepackage{multirow}
\usepackage{hhline}

%% file: commands.tex
\newcommand{\googleai}{Google AI}
\newcommand{\simulamet}{SimulaMet, Norway}
\newcommand{\tum}{Technical University of Munich, Germany}
\newcommand{\uib}{University of Bergen, Norway}
\newcommand{\hkris}{Kristiania University College, Norway}
\newcommand{\triplet}{<Image, Caption1, Caption2> triplet}
\newcommand{\triplets}{<Image, Caption1, Caption2> triplets}
\newcommand{\yt}{YouTube}

\definecolor{ForestGreen}{RGB}{34,139,34}
\definecolor{brickred}{rgb}{0.8, 0.25, 0.33}
\newcommand{\green}[1]{\textcolor{ForestGreen}{#1}}
\newcommand{\red}[1]{\textcolor{brickred}{#1}}

%% file: authors.tex
\author{Shivangi Aneja}
\affiliation{\institution{\tum}}
\author{Cise Midoglu}
\affiliation{\institution{\simulamet}}
\author{Duc-Tien Dang-Nguyen}
\affiliation{\institution{\uib}}
\affiliation{\institution{\hkris}}
\author{Michael Alexander Riegler}
\affiliation{\institution{\simulamet}}
\author{P{\aa}l Halvorsen}
\affiliation{\institution{\simulamet}}
\author{Matthias Nie{\ss}ner}
\affiliation{\institution{\tum}}
\author{Balu Adsumilli}
\affiliation{\institution{\yt}}
\author{Chris Bregler}
\affiliation{\institution{\googleai}}

\renewcommand{\shortauthors}{Aneja et al.}

%% file: sections/0-abstract.tex
\begin{abstract}
\textit{Cheapfake} is a recently coined term that encompasses non-AI (``cheap") manipulations of multimedia content. Cheapfakes are known to be more prevalent than deepfakes. Cheapfake media can be created using editing software for image/video manipulations, or even without using any software, by simply altering the context of an image/video by sharing the media alongside misleading claims. This alteration of context is referred to as \textit{\ac{OOC} misuse} of media. \ac{OOC} media is much harder to detect than fake media, since the images and videos are not tampered. In this challenge, we focus on detecting \ac{OOC} images, and more specifically the misuse of real photographs with conflicting image captions in news items. The aim of this challenge is to develop and benchmark models that can be used to detect whether given samples (news image and associated captions) are \ac{OOC}, based on the recently compiled COSMOS dataset.
\end{abstract}

\keywords{Cheapfakes, Misinformation, News, Out-of-context misuse, Re-contextualized media}

%% file: figures/teaser-figure.tex
\begin{teaserfigure}
  \includegraphics[width=\textwidth]{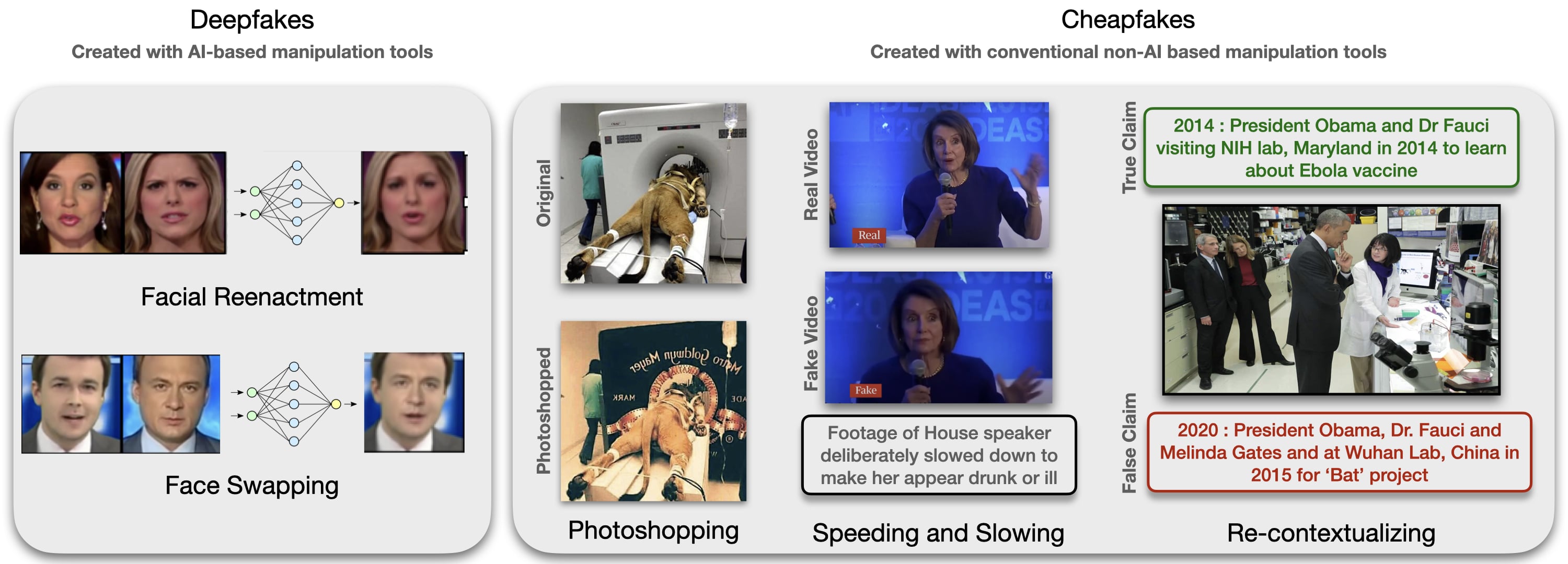}
  \caption{Multimedia manipulation can be broadly grouped into two categories. 
  Deepfakes (left): These are falsified media created using sophisticated AI-based media manipulation tools and techniques. Deepfake videos include re-enactment based techniques where head-pose/expressions are transferred from one video to the other, and swapping methods where facial region is swapped from one video to the other.  
  Cheapfakes (right): These include falsified media created with/without contemporary non-AI based editing tools which are easily accessible. Photoshopping tools can be used to tamper with images. Videos can be sped up or slowed down to change the intent or misrepresent the person in the video. Re-contextualizing includes associating falsified or unrelated claims with a genuine image to misrepresent events or persons. This challenge is focused on detecting re-contextualized cheapfakes.
  Image sources:~\cite{ff_dataset,photoshopped_lion,pelosi_fake,obama_maryland,obama_wuhan}}
  \label{figure:teaser}
\end{teaserfigure}

%% file: sections/1-introduction.tex
\section{Introduction}\label{section:introduction}

The last decade has seen a surge in the use of social media platforms as a means of consuming news. 
In a recent study~\cite{forbes_report}, Forbes reports that social media giant Facebook leads this trend with $36\%$ of its customers using the platform for consuming news. Social media platforms come with a freedom for users to upload and share posts, which has led to the proliferation of fake media on these platforms.

Fake media (including audio, images, videos, and text) circulated on social media platforms can be broadly grouped into two major categories: \textit{deepfakes} and \textit{cheapfakes}, as shown in Figure~\ref{figure:teaser}. Deepfakes are falsified media, most commonly facial videos created using sophisticated AI-based media manipulation tools and techniques. Several deepfake detection methods~\cite{ff_dataset, Nguyen_2019, face_warping, deepfake_inconsistent_head_pose, cozzolino2018forensictransfer,mesonet, agarwal_protecting_2019, li2020face, verdoliva2020media, aneja2020generalized, cozzolino2020idreveal} are in place to monitor and regulate the spread of deepfake videos.

Cheapfake is a general term that encompasses many non-AI (``cheap") manipulations of multimedia content, created without using deep learning methods. Cheapfakes are created with or without contemporary editing tools which are non-AI based and are easily accessible. 
As shown in Figure~\ref{figure:teaser}, cheapfake generation methods can include the use of editing software such as Adobe Photoshop or PremierePro for image manipulations, deliberate alteration of context in news captions, and speeding/slowing of videos. 
We refer readers to the report by Paris \textit{et al.}~\cite{paris2019} for an overview of different types of cheapfakes surfacing the Internet. In fact, certain studies have found cheapfakes to be more prevalent than deepfakes~\cite{factsheet-covid19, mit_tech_report}. 

Depending on the type of cheapfake media, different detection tools need to be developed. For instance, methods to detect image manipulations such as photoshopping and image splicing have been investigated~\cite{Chen2017ImageSD, Cozzolino2015SplicebusterAN, huh2018fighting, wang2019detecting}. Re-contextualization or \acf{OOC} misuse, which include associating falsified or unrelated claims with a genuine image in order to misrepresent events or persons is, however, relatively niche and unexplored. Very recently, Aneja \textit{et al.}~\cite{aneja2021cosmos} introduced this task, provided a dataset of real-world news posts called COSMOS, and proposed a method for detecting cheapfakes which was benchmarked using this dataset. 

Note that \ac{OOC} misuse of images should not be confused with multi-modal (image, caption) fake news detection methods~\cite{zhiwei_17,fauxbuster_2018, Shang2020FauxWardAG, Wang_eann_18,zlatkova-etal-2019-fact,Khattar2019MVAEMV}, which aim to identify fake news where images could be photoshopped and the real counterpart does not even exists. In the case of \ac{OOC} misuse, the images are genuine and the real counterpart always exists. This is illustrated in Figure~\ref{figure:fake-media}.

It is difficult to construct large-scale supervised dataset for the task of detecting \ac{OOC} misuse, due to the relatively small number of images used to spread misinformation compared to the overall number of images shared on the Internet every day. 
To foster research in this direction, automatic methods to create synthetic \ac{OOC} media can be used~\cite{luo2021newsclippings}.

\input{figures/fake-media}

In this challenge, we focus on detecting a specific type of cheapfake -- the alteration of context for news captions, where the key idea is to take an existing genuine image and create a highly convincing but potentially misleading message in order to spread misinformation. 
The challenge is based on the recently compiled COSMOS dataset~\cite{aneja2021cosmos}, tailored specifically to the detection of \ac{OOC} misuse of images. The algorithms developed by participants for this challenge should be able to detect \ac{OOC} \triplets\ in news items. The task is further explained in Section~\ref{section:task} and illustrated in Figure~\ref{figure:task-figure}. 

This article provides basic information about the dataset (Section~\ref{section:dataset}), the task (Section~\ref{section:task}), the evaluation criteria (Section~\ref{section:evaluation}), and the administrative details (Section~\ref{section:administrative}) for the challenge. A challenge webpage~\footnote{\url{https://2021.acmmmsys.org/cheapfake\_challenge.php}\label{footnote:challenge-webpage}} has also been setup for information and official announcements.

%% file: figures/fake-media.tex
\begin{figure}[ht!]
    \centering
    \includegraphics[width=\linewidth]{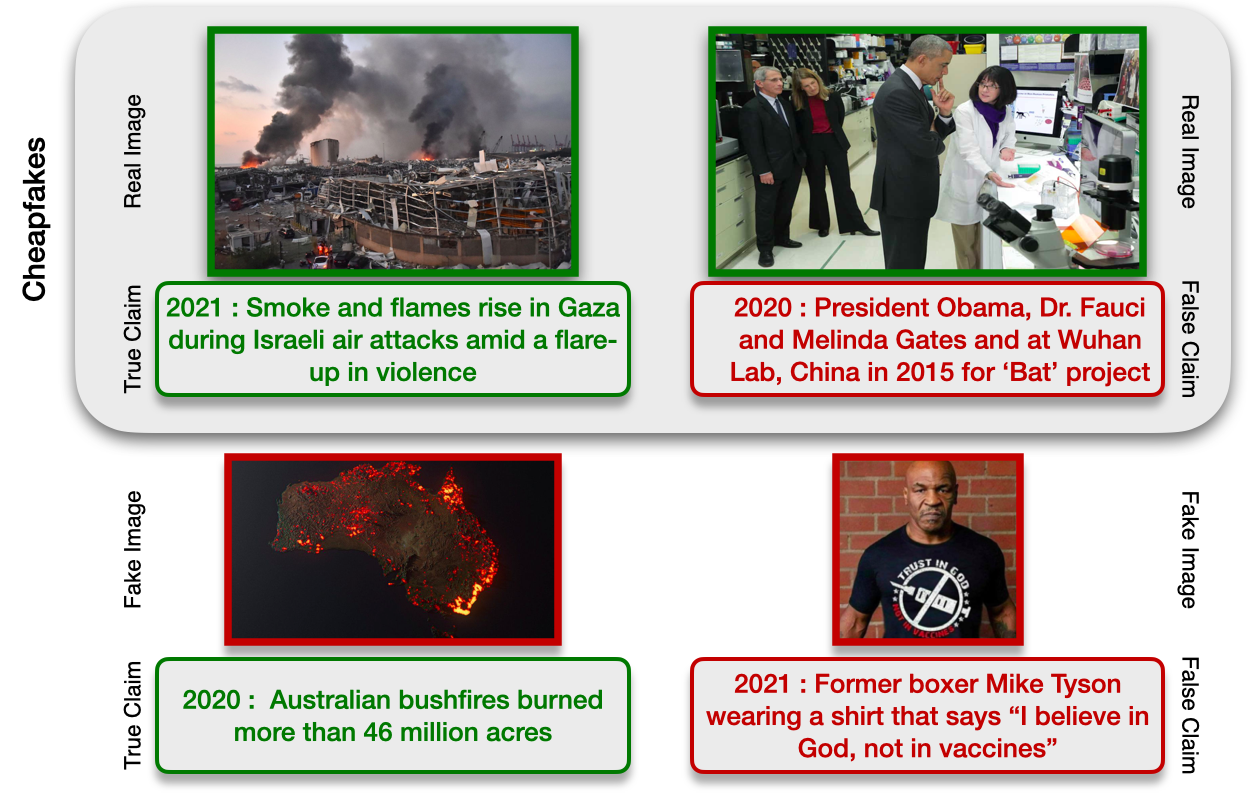}
    \caption{Fauxtography on social media: \green{Green} denotes real image/true claim and \red{Red} denotes fake image/false claim. Different real/fake image and true/false claim combinations used in news posts to spread misleading content can be considered under the umbrella term of \textit{fake news}. Note that here we refer to the misuse of genuine images as \acf{OOC} cheapfakes, such as the use of contextually similar but irrelevant photographs (e.g., different time, different location, different event) with true/false claims, as shown in the top row. 
    Image sources:~\cite{gaza_blast, beirut_blast, obama_wuhan, australia_fire, tyson_vaccine}}
    \label{figure:fake-media}
\end{figure}

%% file: sections/2-dataset.tex
\section{Dataset}\label{section:dataset}

Aneja et al.~\cite{aneja2021cosmos} have created COSMOS, a large-scale dataset of around $200K$ images which have been matched with $450K$ textual captions from different news websites, blogs, and social media posts. Figure~\ref{figure:dataset-distribution} presents the category distribution of the images in the dataset, which were collected from a wide-variety of articles with special focus on topics where misinformation spread is prominent.

\input{figures/dataset-distribution}

Overall, the dataset was gathered from two primary sources: news outlets and fact-checking websites. First, using publicly available news channel APIs, the authors scraped images along with the corresponding captions. Then, reverse-searching these images using Google’s Cloud Vision API, the authors found different contexts across the web in which the same images were shared. Thus, they have obtained several captions per image with varying context (2-4 captions per image).

\input{tables/dataset-statistics}

For this challenge, a part of the COSMOS dataset is sampled and assigned as the public dataset. The public dataset, consisting of the \textit{training}, \textit{validation}\ and \textit{public test} splits, is provided openly to participants for training and testing their algorithms\footnote{Prospective participants are able to get access to the dataset by filling out a Google Form indicated on the challenge webpage.\label{footnote:dataset-form}}. The remaining part of the COSMOS dataset is augmented with new samples and modified to create the \textit{hidden test} split, which is not made publicly available, and will be used by the challenge organizers to evaluate the submissions. Table~\ref{table:dataset-statistics} provides statistics about the overall challenge dataset. 

\subsection{Training and Validation Splits}

The training and validation splits are provided as \ac{JSON} formatted text files called \texttt{train.json} and \texttt{val.json}, where each data sample is stored as a dictionary (see Listing~\ref{lst:train-val}). 
The attributes in \texttt{train.json} and \texttt{val.json} are as follows.

\begin{itemize}
    \item \texttt{img\_local\_path}: Source path for the image in the dataset directory.
    \item \texttt{articles}: List of dictionaries containing metadata for every caption associated with the image.
    \item \texttt{caption:} Original caption scraped from the news website.
    \item \texttt{article\_url}: Link to the website from where the image and caption were scraped.
    \item \texttt{caption\_modified}: Modified \texttt{caption} after applying Spacy NER~\footnote{https://spacy.io/api/entityrecognizer}. Authors in~\cite{aneja2021cosmos} use this caption as an input to their model during experiments.
    \item \texttt{entity\_list}: List of mappings between the modified named entities in \texttt{caption} with the corresponding hypernyms.
    \item \texttt{maskrcnn\_bboxes}: List of detected bounding boxes corresponding to the image. (x1,y1) refers to the start vertex of the rectangle and (x2, y2) refers to end vertex of the rectangle. Note that for detecting bounding boxes, the authors in~\cite{aneja2021cosmos} use the Detectron2 pretrained model~\footnote{\url{https://github.com/facebookresearch/detectron2/blob/master/MODEL\_ZOO.md}} available under~\footnote{\url{https://github.com/facebookresearch/detectron2/blob/master/configs/COCO-Keypoints/keypoint\_rcnn\_X\_101\_32x8d\_FPN\_3x.yaml}}. They detect up to $10$ bounding boxes per image. 
\end{itemize}

\begin{lstlisting}[caption={File structure for train.json and val.json},label={lst:train-val},float,floatplacement=H]
    {
    "img_local_path": <img_path>, 
    "articles": [
        {"caption": <caption1>,
        "article_url": <url1>,
        "caption_modified": <caption_mod1>,
        "entity_list": <entity_list1>},
        {"caption": <caption2>,
        "article_url": <url2>,
        "caption_modified": <caption_mod2>,
        "entity_list": <entity_list2>},
        {"caption": <caption3>,
        "article_url": <url3>,
        "caption_modified": <caption_mod3>,
        "entity_list": <entity_list3>},
        ... ],
    "maskrcnn_bboxes": [
        [x1,y1,x2,y2], 
        [x1,y1,x2,y2], 
        ... ]
    }
\end{lstlisting}

\subsection{Test Splits}

The public test split is provided as a \ac{JSON} formatted text file called \texttt{test.json}, and has the structure shown in Listing~\ref{lst:test}. The hidden test split is structurally identical to the public test split. 
The attributes in \texttt{test.json} are as follows.

\begin{itemize}
    \item \texttt{img\_local\_path}: Source path for the image in the dataset directory.
    \item \texttt{caption1}: First caption associated with the image.
    \item \texttt{caption1\_modified}: Modified \texttt{caption1} after applying Spacy NER.
    \item \texttt{caption1\_entities}: List of mappings between the modified named entities in \texttt{caption1} with the corresponding hypernyms.
    \item \texttt{caption2}: Second caption associated with the image.
    \item \texttt{caption2\_modified}: Modified \texttt{caption2} after applying Spacy NER.
    \item \texttt{caption2\_entities}: List of mappings between the modified named entities in \texttt{caption2} with the corresponding hypernyms.
    \item \texttt{article\_url}: Link to the website from where the image and caption were scraped.
    \item \texttt{label}: Class label indicating whether the two captions are out-of-context with respect to the image (1=\acf{OOC}, 0=\acf{NOOC})
    \item \texttt{maskrcnn\_bboxes}: List of detected bounding boxes corresponding to the image. (x1,y1) refers to the start vertex of the rectangle and (x2, y2) refers to the end vertex of the rectangle.
\end{itemize}

\begin{lstlisting}[caption={File structure for test.json},label={lst:test},float,floatplacement=H]
    {	
    "img_local_path": <img_path>,
    "caption1": <caption1>,
    "caption1_modified": <caption1_modified>,
    "caption1_entities": <caption1_entities>,
    "caption2": <caption2>,
    "caption2_modified": <caption2_modified>,
    "caption2_entities": <caption2_entities>,
    "article_url": <article_url>,
    "label": "ooc/not-ooc",
    "maskrcnn_bboxes": [
        [x1,y1,x2,y2], 
        [x1,y1,x2,y2], 
        ... ]
    }
\end{lstlisting}

%% file: figures/dataset-distribution.tex
\begin{figure}[h!]
    \centering
    \includegraphics[width=\linewidth]{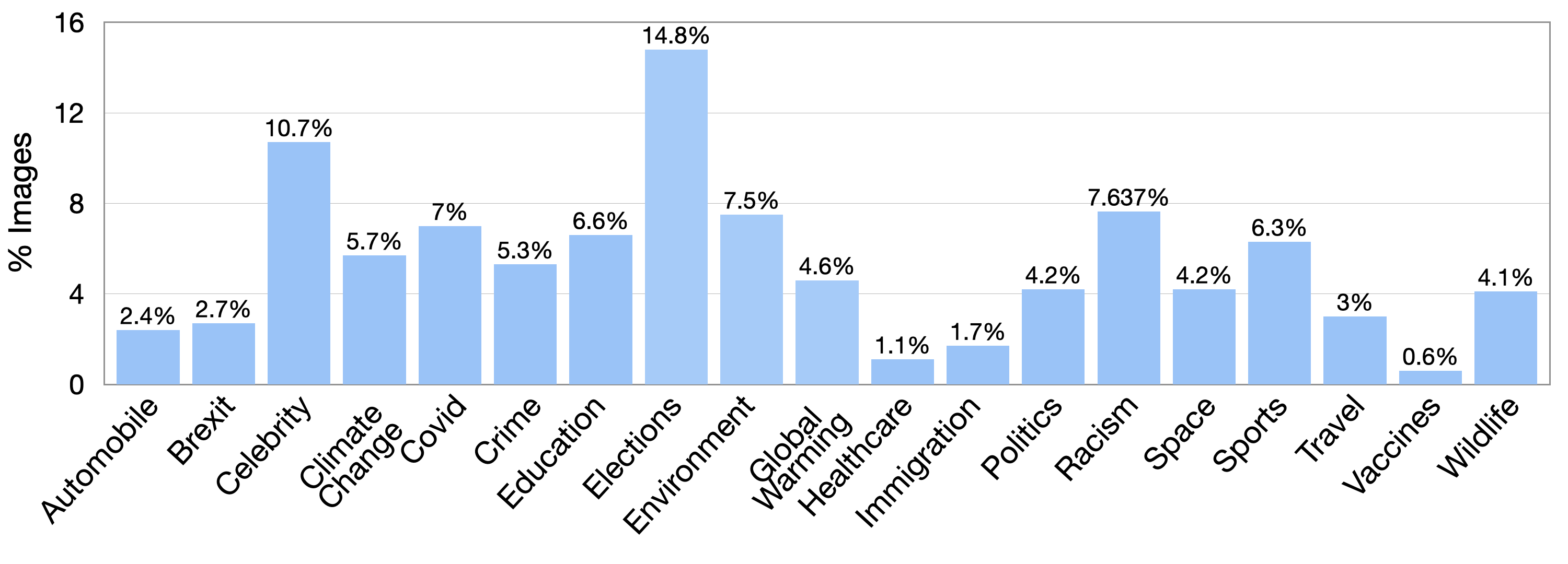}
    \caption{Distribution of the images in the COSMOS dataset per category.}
    \label{figure:dataset-distribution}
\end{figure}

%% file: tables/dataset-statistics.tex
\begin{table}[h!]
    \centering
    \begin{tabular}{|c|r|r|c|}
    \hline
    \textbf{Split} & \textbf{\#Images} & \textbf{\#Captions} & \textbf{Context Annotation} \\
    \hline
    Training & $161752$ & $360749$ & No \\
    \hline
    Validation & $41006$ & $90036$ & No \\
    \hline
    Public Test & $1000$ & $2000$ & Yes\\
    \hline
    Hidden Test & \texttt{N/A} & \texttt{N/A} & Yes\\
    \hline
    \end{tabular}
    \caption{Challenge dataset statistics.}
    \label{table:dataset-statistics}
\end{table}

%% file: sections/3-task.tex
\section{Task Description}\label{section:task}

This challenge invites participants to develop a model using the dataset provided by the organizers, for the detection of \ac{OOC} image captions that might be accompanying news images. 
For each sample in a given test split described above, their model must detect whether the \triplet\ is \ac{OOC} or \ac{NOOC}, and output the corresponding class label: 1=\ac{OOC} or 0=\ac{NOOC}. 

\subsection{Considerations}

There are two considerations in the fulfilment of the above described task.

\begin{itemize}
    \item \textbf{Binary detection performance:} The first goal of the task is to achieve high detection performance, i.e., to be able to detect whether \triplets\ are \ac{OOC} or \ac{NOOC}, successfully. This speaks to \textit{effectiveness}. Participant models are evaluated based on the \acf{E1} described in Section~\ref{section:evaluation}. 
    
    \item \textbf{Latency and complexity:} In certain scenarios, having an idea about the potential misuse of images in real-time and with minimal resources can be more important than the detection performance itself. This speaks to \textit{efficiency}. We take this aspect into consideration by introducing an additional goal: having low latency and low complexity. Participant models are evaluated based on the \acf{E2} described in Section~\ref{section:evaluation}. 
\end{itemize}

\input{figures/task-figure}

\subsection{Baseline Model}\label{section:baseline-model}

Prospective participants are provided with the pre-trained model from~\cite{aneja2021cosmos} upon request. This model is provided only as a reference, where interested participants are encouraged to use the model for reproducing the results from the original paper, and/or as a baseline for developing their own models for this challenge. 

The core idea of~\cite{aneja2021cosmos} is a self-supervised training strategy where only captioned images are needed: no explicit \ac{OOC} annotations are required during training, which could be potentially difficult to acquire in large numbers. The image captions from the dataset are established as matches, and random captions from other images as non-matches. Using these matches vs. non-matches as a loss function, the authors are able to learn the co-occurrence patterns of images with textual descriptions in order to determine whether an image appears to be \ac{OOC} with respect to textual claims. During training, their method only learns to selectively align individual objects in an image with textual claims, without explicit \ac{OOC} supervision. At test time, they correlate these alignment predictions between the two captions for the input image. If both texts correspond to the same object but their meaning is semantically different, they infer that the image is used \ac{OOC}.

\subsection{Test Environment}

Participants are free to develop their models in any language or platform they prefer. However, a submission in the form of a Docker image is required for evaluation. This image should include all the required dependencies and should be possible to run using the latest version of Docker (releases for Linux/Mac/Windows are available under~\footnote{\url{https://docs.docker.com/get-docker/}}). Note that data should not be included within the Docker image itself, as it will be injected during evaluation. Participants can assume that the test dataset will be located at \texttt{/mmsys21cheapfakes}. Sample Docker file instructions can be found in the official GitHub repository for the challenge~\cite{github-challenge}.

%% file: figures/task-figure.tex
\begin{figure}[ht!]
    \centering
    \includegraphics[width=\linewidth]{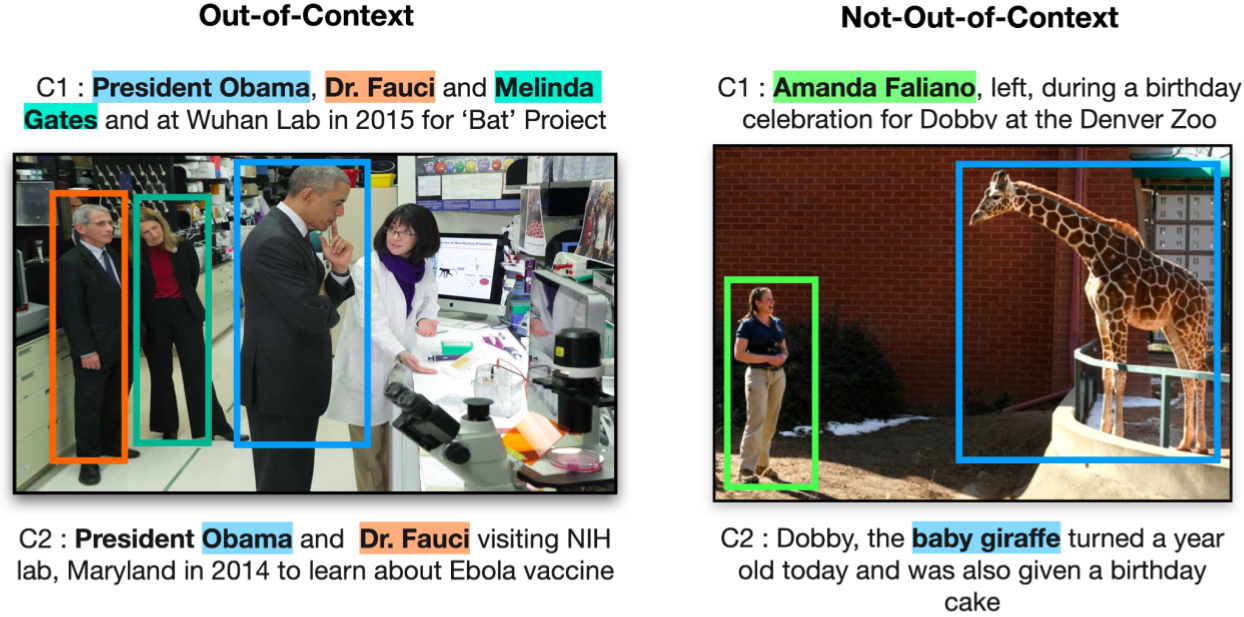}
    \caption{Task description: Each image is accompanied by two captions it was shared together with on the Internet. On the left, one of the two captions is misleading with an alteration of context, hence \acf{OOC}. On the right, none of the two captions are misleading, hence \acf{NOOC}. Given \triplets\ as input, the model should predict corresponding class labels. Image source:~\cite{aneja2021cosmos}}
    \label{figure:task-figure}
\end{figure}

%% file: sections/4-evaluation.tex
\input{figures/timeline}

\section{Evaluation Criteria}\label{section:evaluation}

In order to rank participant models, two aggregate scores will be used: \ac{E1} and \ac{E2}.

\subsection{\acf{E1}}

Given the following definitions 
\begin{itemize}
\item \textbf{\ac{TP}:} Number of samples correctly identified as \ac{OOC}
\item \textbf{\ac{TN}:} Number of samples correctly identified as \ac{NOOC}
\item \textbf{\ac{FP}:} Number of samples incorrectly identified as \ac{OOC}
\item \textbf{\ac{FN}:} Number of samples incorrectly identified as \ac{NOOC}
\end{itemize}

\noindent The \textit{effectiveness} of participant models will be evaluated according to the following $5$ metrics: accuracy, precision, recall, F1-score~\cite{f1-score}, and \ac{MCC}~\cite{mcc}. 
Authors are asked to calculate the $5$ metrics for their model and include these values in their manuscript. E1 will be a function of accuracy, F1-score and \ac{MCC}, to be calculated by the organizers.

\begin{equation}
\label{equation:accuracy}
    Accuracy = \frac{TP + TN}{TP + FP + TN + FN}
\end{equation}

\begin{equation}
\label{equation:precision}
    Precision = \frac{TP}{TP + FP}
\end{equation}

\begin{equation}
\label{equation:recall}
    Recall = \frac{TP}{TP + FN}
\end{equation}

\begin{equation}
\label{equation:f1-score}
\begin{split}
F1 & = 2 \frac{(Recall \times Precision)}{(Recall + Precision)} \\
& = \frac{TP}{TP + \frac{1}{2} (FP + TP)}
\end{split}
\end{equation}

\begin{equation}
\label{equation:mcc}
MCC = \frac{(TP \times TN) - (FP \times FN)}
{\sqrt{(TP + FP)(TP + FN)(TN + FP)(TN + FN)}}
\end{equation}

\subsection{\acf{E2}}

The \textit{efficiency} of participant models will be evaluated according to the following $3$ metrics: latency, number of parameters, and model size. Participants are asked to calculate the $3$ metrics for their model and include these values in their manuscript. E2 will be a function of latency, complexity-1 and complexity-2, to be calculated by the organizers. 

\begin{itemize}
    \item Latency: Average runtime per sample\footnote{Arithmetic mean of the runtime per sample, calculated over all samples in the public test split.} (ms)
    \item Complexity-1: Number of trainable parameters in the model (million)
    \item Complexity-2: Model size (MB)
\end{itemize}

\subsection{Baseline Model Scores}

As a reference, we run the baseline model described in Section~\ref{section:baseline-model} on the public test split described in Section~\ref{section:dataset}. Results can be found in Table~\ref{table:baseline-scores}.

\input{tables/baseline-scores}

This model achieves $82\%$ accuracy when run on the public test split. Note that the accuracy value is different from the original paper~\cite{aneja2021cosmos}, where the model is run on a slightly different dataset.

%% file: figures/timeline.tex
\begin{figure*}[ht!]
    \centering
    \includegraphics[width=0.95\linewidth]{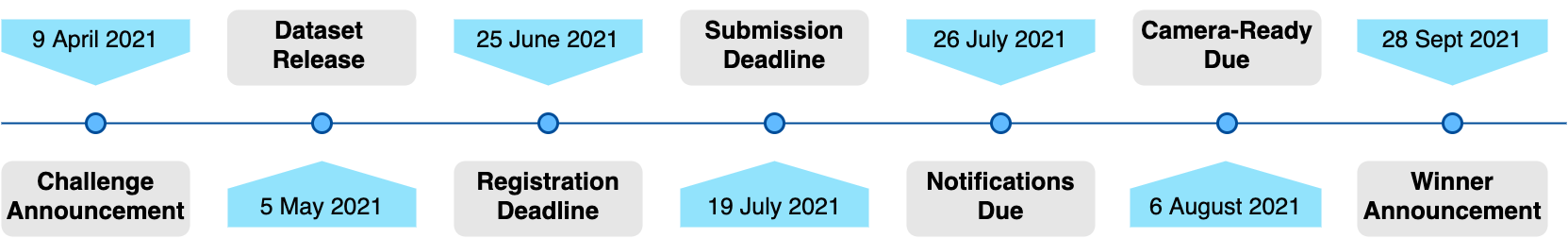}
    \caption{Timeline for the MMSys'21 Grand Challenge on Detecting Cheapfakes.}
    \label{figure:timeline}
\end{figure*}

%% file: tables/baseline-scores.tex
\begin{table}[h!]
    \centering
    \begin{tabular}{|c|l|c|}
    \hhline{~--}
    \multicolumn{1}{c}{} 
    & \multicolumn{1}{|c|}{\textbf{Metric}} & \textbf{Value} \\
    \hline
    \multirow{5}{*}{E1} & Accuracy (Eq.~\ref{equation:accuracy}) & $0.82$  \\
    \hhline{~--}
     & Precision (Eq.~\ref{equation:precision}) & $0.85$ \\
    \hhline{~--}
     & Recall (Eq.~\ref{equation:recall}) & $0.77$\\
    \hhline{~--}
     & F1-score (Eq.~\ref{equation:f1-score}) & $0.81$\\
    \hhline{~--}
     & \ac{MCC} (Eq.~\ref{equation:mcc}) & $0.64$ \\
    \hline
    \multirow{3}{*}{E2} & Latency: runtime for one data sample (ms) & $35$ \\
    \hhline{~--}
    & Complexity-1: \# trainable parameters (million) & $2.5$ \\
    \hhline{~--}
    & Complexity-2: model size (MB) & $10.2$ \\
    \hline
    \end{tabular}
    \caption{Baseline model results on the public test split.}
    \label{table:baseline-scores}
\end{table}

%% file: sections/5-administrative.tex
\section{Administrative Details}\label{section:administrative}

\subsection{Timeline and Awards}

The timeline for the grand challenge is indicated in Figure~\ref{figure:timeline}. The challenge was announced on April 9, and the dataset was released\textsuperscript{\ref{footnote:dataset-form}} on May 5. 
The registration/submission site\footnote{\url{https://mmsys2021challenges.hotcrp.com/}\label{footnote:submission-site}} was opened on May 31. 
The registration and submission deadlines are June 25 and July 19, respectively. 
The authors will be notified after a review process (acceptance notifications are due July 26). Authors of accepted papers need to prepare a camera-ready version according to the instructions provided below in Section~\ref{section:submission-instructions}, in order for their papers to be published in the ACM Digital Library. Camera-ready manuscripts are due August 6. 

The challenge is open to any individual, commercial or academic institution. Winners will be chosen by a committee appointed by the challenge organizers and the decision will be final. The results will be announced during the ACM Multimedia Systems Conference (MMSys'21)\footnote{\url{https://2021.acmmmsys.org/}} which will take place between September 28 and October 1. The winner will be awarded $5,000$ USD and the runner-up will be awarded $2,500$ USD. If contributions of sufficient quality are not received, then some or all of the awards may not be granted.

\subsection{Submission Instructions}\label{section:submission-instructions}

Submissions should be prepared according to the guidelines provided below.

\begin{itemize}
    \item The manuscript should provide enough details for the implemented algorithm in a short technical paper, and include references to the public repository for the source code and the Docker image.
    \item Page count: Up to $6$ pages plus an optional page for references only.
    \item Style: Single blind using the ACM proceedings template\footnote{\url{https://www.acm.org/publications/proceedings-template}}. 
    \item ACM header: ``MMSys'21, Sept. 28-Oct. 1, 2021, Istanbul, Turkey" on the left, and author names (on even pages) / title (on odd pages, except the first page) on the right.
    \item Format: \ac{PDF}
    \item Online submission: \url{https://mmsys2021challenges.hotcrp.com/}
\end{itemize}

\subsection{Participant Support}

Along with the official webpage for the challenge\textsuperscript{\ref{footnote:challenge-webpage}}, a Google Group\footnote{"MMSys'21 Grand Challenge on Detecting Cheapfakes" group homepage: \url{https://groups.google.com/g/mmsys21-grandchallenge-cheapfakes/}, e-mail: \href{mailto:mmsys21-grandchallenge-cheapfakes@googlegroups.com}{mmsys21-grandchallenge-cheapfakes@googlegroups.com}.} and a Slack channel\footnote{ACM MMSys Slack workspace: \url{https://tinyurl.com/mmsys-slack/}, channel for grand challenge: \#2021-gc-cheapfakes} have been established to support prospective participants. Interested participants can find the previously asked questions and join interactive discussions on these platforms. 

%% file: sections/6-conclusion.tex
\section{Conclusion and Outlook}\label{section:conclusion}

The MMSys'21 Grand Challenge on Detecting Cheapfakes addresses the relatively new but prominent problem of \textit{cheapfake} multimedia content in news. More specifically, it focuses on the \ac{OOC} misuse of images in news items. As emphasized in the Section~\ref{section:introduction}, this is a relatively novel area of research, in comparison to deepfakes, and is distinguishable from the wider field of fake news in the following way: the term ``fake news" traditionally refers to the use of either \textit{fake} multimedia content or \textit{false} captions, whereas \ac{OOC} misuse refers to the scenario where the multimedia content in the news item (images in this case) is decidedly not fake, and the captions may or may not be false. Rather, the misinformation results from the \ac{OOC} combination of the two.

With this challenge, we firstly aim to motivate researchers to develop different methods for addressing this particular problem, and to benchmark a number of proposed models for detecting the \ac{OOC} misuse of photographs in news items. 
Secondly, through the dissection and analysis of the COSMOS dataset, we aim to encourage the in-depth understanding and later generation of (further) supervised datasets for this task. Algorithmic benchmarking is an efficient approach to analyze the results from different detection methods, but on top of evaluating the models themselves, the comparison of different approaches can help identify the potentials as well as the shortcomings of existing open datasets. 

There is a growing interest from the scientific community towards addressing the problem of misinformation in general, and towards the detection of deepfakes in particular. We hope that this challenge can increase awareness regarding the prominence of cheapfakes as well, and that in the future, the methods presented within this context can evolve into systems that support researchers, regulatory bodies, news consumers, and the general public in their search for a safe and truthful information ecosystem.

%% file: main.bbl

\begin{thebibliography}{38}


\ifx \showCODEN    \undefined \def \showCODEN     #1{\unskip}     \fi
\ifx \showDOI      \undefined \def \showDOI       #1{#1}\fi
\ifx \showISBNx    \undefined \def \showISBNx     #1{\unskip}     \fi
\ifx \showISBNxiii \undefined \def \showISBNxiii  #1{\unskip}     \fi
\ifx \showISSN     \undefined \def \showISSN      #1{\unskip}     \fi
\ifx \showLCCN     \undefined \def \showLCCN      #1{\unskip}     \fi
\ifx \shownote     \undefined \def \shownote      #1{#1}          \fi
\ifx \showarticletitle \undefined \def \showarticletitle #1{#1}   \fi
\ifx \showURL      \undefined \def \showURL       {\relax}        \fi
\providecommand\bibfield[2]{#2}
\providecommand\bibinfo[2]{#2}
\providecommand\natexlab[1]{#1}
\providecommand\showeprint[2][]{arXiv:#2}

\bibitem[\protect\citeauthoryear{??}{pel}{2019}]%
        {pelosi_fake}
 \bibinfo{year}{2019}\natexlab{}.
\newblock \bibinfo{booktitle}{\emph{Real vs fake: debunking the 'drunk' Nancy
  Pelosi footage}}.
\newblock
\urldef\tempurl%
\url{https://www.theguardian.com/technology/2019/may/24/facebook-leaves-fake-nancy-pelosi-video-on-site}
\showURL{%
\tempurl}


\bibitem[\protect\citeauthoryear{Afchar, Nozick, Yamagishi, and Echizen}{Afchar
  et~al\mbox{.}}{2018}]%
        {mesonet}
\bibfield{author}{\bibinfo{person}{Darius Afchar}, \bibinfo{person}{Vincent
  Nozick}, \bibinfo{person}{Junichi Yamagishi}, {and} \bibinfo{person}{Isao
  Echizen}.} \bibinfo{year}{2018}\natexlab{}.
\newblock \bibinfo{title}{MesoNet: a Compact Facial Video Forgery Detection
  Network}.
\newblock
\newblock
\showISBNx{9781538665367}
\urldef\tempurl%
\url{https://doi.org/10.1109/wifs.2018.8630761}
\showDOI{\tempurl}


\bibitem[\protect\citeauthoryear{Agarwal, Farid, Gu, He, Nagano, and
  Li}{Agarwal et~al\mbox{.}}{2019}]%
        {agarwal_protecting_2019}
\bibfield{author}{\bibinfo{person}{S. Agarwal}, \bibinfo{person}{H. Farid},
  \bibinfo{person}{Yuming Gu}, \bibinfo{person}{Mingming He},
  \bibinfo{person}{Koki Nagano}, {and} \bibinfo{person}{H. Li}.}
  \bibinfo{year}{2019}\natexlab{}.
\newblock \showarticletitle{Protecting World Leaders Against Deep Fakes}. In
  \bibinfo{booktitle}{\emph{CVPR Workshops}}.
\newblock


\bibitem[\protect\citeauthoryear{Aljazeera}{Aljazeera}{2021}]%
        {gaza_blast}
\bibfield{author}{\bibinfo{person}{Aljazeera}.}
  \bibinfo{year}{2021}\natexlab{}.
\newblock \bibinfo{booktitle}{\emph{What led to the most recent
  Israel-Palestine escalation?}}
\newblock
\urldef\tempurl%
\url{https://www.aljazeera.com/news/2021/5/12/what-lead-up-to-most-recent-israel-palestine-escalation}
\showURL{%
Retrieved May 12, 2021 from \tempurl}


\bibitem[\protect\citeauthoryear{Aneja, Bregler, and Nießner}{Aneja
  et~al\mbox{.}}{2021}]%
        {aneja2021cosmos}
\bibfield{author}{\bibinfo{person}{Shivangi Aneja}, \bibinfo{person}{Chris
  Bregler}, {and} \bibinfo{person}{Matthias Nießner}.}
  \bibinfo{year}{2021}\natexlab{}.
\newblock \bibinfo{title}{{COSMOS}: Catching Out-of-Context Misinformation with
  Self-Supervised Learning}.
\newblock
\newblock
\showeprint[arxiv]{2101.06278}~[cs.CV]


\bibitem[\protect\citeauthoryear{Aneja and Nießner}{Aneja and
  Nießner}{2020}]%
        {aneja2020generalized}
\bibfield{author}{\bibinfo{person}{Shivangi Aneja} {and}
  \bibinfo{person}{Matthias Nießner}.} \bibinfo{year}{2020}\natexlab{}.
\newblock \bibinfo{title}{Generalized Zero and Few-Shot Transfer for Facial
  Forgery Detection}.
\newblock
\newblock
\showeprint[arxiv]{2006.11863}~[cs.CV]


\bibitem[\protect\citeauthoryear{Boredpanda}{Boredpanda}{2019}]%
        {photoshopped_lion}
\bibfield{author}{\bibinfo{person}{Boredpanda}.}
  \bibinfo{year}{2019}\natexlab{}.
\newblock \bibinfo{booktitle}{\emph{30 Fake Viral Photos People Believed Were
  Real}}.
\newblock
\urldef\tempurl%
\url{https://www.boredpanda.com/fake-news-photos-viral-photoshop/}
\showURL{%
\tempurl}


\bibitem[\protect\citeauthoryear{Brennen, Simon, Howard, and Nielsen}{Brennen
  et~al\mbox{.}}{2020}]%
        {factsheet-covid19}
\bibfield{author}{\bibinfo{person}{J.~Scott Brennen}, \bibinfo{person}{Felix~M.
  Simon}, \bibinfo{person}{Philip~N. Howard}, {and}
  \bibinfo{person}{Rasmus~Kleis Nielsen}.} \bibinfo{year}{2020}\natexlab{}.
\newblock \bibinfo{booktitle}{\emph{Types, Sources, and Claims of {COVID-19}
  Misinformation}}.
\newblock
\urldef\tempurl%
\url{http://www.primaonline.it/wp-content/uploads/2020/04/COVID-19\_reuters.pdf}
\showURL{%
\tempurl}


\bibitem[\protect\citeauthoryear{Chen, McCloskey, and Yu}{Chen
  et~al\mbox{.}}{2017}]%
        {Chen2017ImageSD}
\bibfield{author}{\bibinfo{person}{Can Chen}, \bibinfo{person}{Scott
  McCloskey}, {and} \bibinfo{person}{J. Yu}.} \bibinfo{year}{2017}\natexlab{}.
\newblock \showarticletitle{Image Splicing Detection via Camera Response
  Function Analysis}.
\newblock \bibinfo{journal}{\emph{2017 IEEE Conference on Computer Vision and
  Pattern Recognition (CVPR)}} (\bibinfo{year}{2017}),
  \bibinfo{pages}{1876--1885}.
\newblock


\bibitem[\protect\citeauthoryear{Cozzolino, Poggi, and Verdoliva}{Cozzolino
  et~al\mbox{.}}{2015}]%
        {Cozzolino2015SplicebusterAN}
\bibfield{author}{\bibinfo{person}{D. Cozzolino}, \bibinfo{person}{G. Poggi},
  {and} \bibinfo{person}{L. Verdoliva}.} \bibinfo{year}{2015}\natexlab{}.
\newblock \showarticletitle{Splicebuster: A new blind image splicing detector}.
\newblock \bibinfo{journal}{\emph{2015 IEEE International Workshop on
  Information Forensics and Security (WIFS)}} (\bibinfo{year}{2015}),
  \bibinfo{pages}{1--6}.
\newblock


\bibitem[\protect\citeauthoryear{Cozzolino, Rössler, Thies, Nießner, and
  Verdoliva}{Cozzolino et~al\mbox{.}}{2020}]%
        {cozzolino2020idreveal}
\bibfield{author}{\bibinfo{person}{Davide Cozzolino}, \bibinfo{person}{Andreas
  Rössler}, \bibinfo{person}{Justus Thies}, \bibinfo{person}{Matthias
  Nießner}, {and} \bibinfo{person}{Luisa Verdoliva}.}
  \bibinfo{year}{2020}\natexlab{}.
\newblock \bibinfo{title}{ID-Reveal: Identity-aware DeepFake Video Detection}.
\newblock
\newblock
\showeprint[arxiv]{2012.02512}~[cs.CV]


\bibitem[\protect\citeauthoryear{Cozzolino, Thies, R{\"o}ssler, Riess,
  Nie{\ss}ner, and Verdoliva}{Cozzolino et~al\mbox{.}}{2018}]%
        {cozzolino2018forensictransfer}
\bibfield{author}{\bibinfo{person}{Davide Cozzolino}, \bibinfo{person}{Justus
  Thies}, \bibinfo{person}{Andreas R{\"o}ssler}, \bibinfo{person}{Christian
  Riess}, \bibinfo{person}{Matthias Nie{\ss}ner}, {and} \bibinfo{person}{Luisa
  Verdoliva}.} \bibinfo{year}{2018}\natexlab{}.
\newblock \showarticletitle{Forensictransfer: Weakly-supervised domain
  adaptation for forgery detection}.
\newblock \bibinfo{journal}{\emph{arXiv preprint arXiv:1812.02510}}
  (\bibinfo{year}{2018}).
\newblock


\bibitem[\protect\citeauthoryear{Evon}{Evon}{2020}]%
        {obama_wuhan}
\bibfield{author}{\bibinfo{person}{Dan Evon}.} \bibinfo{year}{2020}\natexlab{}.
\newblock \bibinfo{booktitle}{\emph{Is This Obama, Fauci, and Gates at a Wuhan
  Lab in 2015?}}
\newblock
\urldef\tempurl%
\url{https://www.snopes.com/fact-check/obama-fauci-gates-wuhan-lab/}
\showURL{%
Retrieved July 13, 2020 from \tempurl}


\bibitem[\protect\citeauthoryear{Evon}{Evon}{2021}]%
        {tyson_vaccine}
\bibfield{author}{\bibinfo{person}{Dan Evon}.} \bibinfo{year}{2021}\natexlab{}.
\newblock \bibinfo{booktitle}{\emph{Mike Tyson’s Anti-Vaccine Shirt Photo Is
  Fake}}.
\newblock
\urldef\tempurl%
\url{https://www.snopes.com/fact-check/mike-tyson-anti-vaccine-shirt/}
\showURL{%
Retrieved July 1, 2021 from \tempurl}


\bibitem[\protect\citeauthoryear{Haverty}{Haverty}{2020}]%
        {beirut_blast}
\bibfield{author}{\bibinfo{person}{Dan Haverty}.}
  \bibinfo{year}{2020}\natexlab{}.
\newblock \bibinfo{booktitle}{\emph{Thousands Injured in Giant Beirut Blast}}.
\newblock
\urldef\tempurl%
\url{https://foreignpolicy.com/2020/08/04/thousands-injured-in-giant-beirut-blast/}
\showURL{%
Retrieved Aug 4, 2020 from \tempurl}


\bibitem[\protect\citeauthoryear{Huh, Liu, Owens, and Efros}{Huh
  et~al\mbox{.}}{2018}]%
        {huh2018fighting}
\bibfield{author}{\bibinfo{person}{Minyoung Huh}, \bibinfo{person}{Andrew Liu},
  \bibinfo{person}{Andrew Owens}, {and} \bibinfo{person}{Alexei~A. Efros}.}
  \bibinfo{year}{2018}\natexlab{}.
\newblock \bibinfo{title}{Fighting Fake News: Image Splice Detection via
  Learned Self-Consistency}.
\newblock
\newblock
\showeprint[arxiv]{1805.04096}~[cs.CV]


\bibitem[\protect\citeauthoryear{Jin, Cao, Guo, Zhang, and Luo}{Jin
  et~al\mbox{.}}{2017}]%
        {zhiwei_17}
\bibfield{author}{\bibinfo{person}{Zhiwei Jin}, \bibinfo{person}{Juan Cao},
  \bibinfo{person}{Han Guo}, \bibinfo{person}{Yongdong Zhang}, {and}
  \bibinfo{person}{Jiebo Luo}.} \bibinfo{year}{2017}\natexlab{}.
\newblock \showarticletitle{Multimodal Fusion with Recurrent Neural Networks
  for Rumor Detection on Microblogs}. In \bibinfo{booktitle}{\emph{Proceedings
  of the 25th ACM International Conference on Multimedia}} (Mountain View,
  California, USA) \emph{(\bibinfo{series}{MM '17})}.
  \bibinfo{publisher}{Association for Computing Machinery},
  \bibinfo{address}{New York, NY, USA}, \bibinfo{pages}{795–816}.
\newblock
\showISBNx{9781450349062}
\urldef\tempurl%
\url{https://doi.org/10.1145/3123266.3123454}
\showDOI{\tempurl}


\bibitem[\protect\citeauthoryear{Khattar, Goud, Gupta, and Varma}{Khattar
  et~al\mbox{.}}{2019}]%
        {Khattar2019MVAEMV}
\bibfield{author}{\bibinfo{person}{Dhruv Khattar},
  \bibinfo{person}{Jaipal~Singh Goud}, \bibinfo{person}{Manish Gupta}, {and}
  \bibinfo{person}{Vasudeva Varma}.} \bibinfo{year}{2019}\natexlab{}.
\newblock \showarticletitle{MVAE: Multimodal Variational Autoencoder for Fake
  News Detection}.
\newblock \bibinfo{journal}{\emph{The World Wide Web Conference}}
  (\bibinfo{year}{2019}).
\newblock


\bibitem[\protect\citeauthoryear{Li, Bao, Zhang, Yang, Chen, Wen, and Guo}{Li
  et~al\mbox{.}}{2020}]%
        {li2020face}
\bibfield{author}{\bibinfo{person}{Lingzhi Li}, \bibinfo{person}{Jianmin Bao},
  \bibinfo{person}{Ting Zhang}, \bibinfo{person}{Hao Yang},
  \bibinfo{person}{Dong Chen}, \bibinfo{person}{Fang Wen}, {and}
  \bibinfo{person}{Baining Guo}.} \bibinfo{year}{2020}\natexlab{}.
\newblock \showarticletitle{Face x-ray for more general face forgery
  detection}. In \bibinfo{booktitle}{\emph{Proceedings of the IEEE/CVF
  Conference on Computer Vision and Pattern Recognition}}.
  \bibinfo{pages}{5001--5010}.
\newblock


\bibitem[\protect\citeauthoryear{Li and Lyu}{Li and Lyu}{2018}]%
        {face_warping}
\bibfield{author}{\bibinfo{person}{Yuezun Li} {and} \bibinfo{person}{Siwei
  Lyu}.} \bibinfo{year}{2018}\natexlab{}.
\newblock \bibinfo{title}{Exposing DeepFake Videos By Detecting Face Warping
  Artifacts}.
\newblock
\newblock
\showeprint[arxiv]{1811.00656}~[cs.CV]


\bibitem[\protect\citeauthoryear{Luo, Darrell, and Rohrbach}{Luo
  et~al\mbox{.}}{2021}]%
        {luo2021newsclippings}
\bibfield{author}{\bibinfo{person}{Grace Luo}, \bibinfo{person}{Trevor
  Darrell}, {and} \bibinfo{person}{Anna Rohrbach}.}
  \bibinfo{year}{2021}\natexlab{}.
\newblock \bibinfo{title}{NewsCLIPpings: Automatic Generation of Out-of-Context
  Multimodal Media}.
\newblock
\newblock
\showeprint[arxiv]{2104.05893}~[cs.CV]


\bibitem[\protect\citeauthoryear{Midoglu and Aneja}{Midoglu and Aneja}{2021}]%
        {github-challenge}
\bibfield{author}{\bibinfo{person}{Cise Midoglu} {and}
  \bibinfo{person}{Shivangi Aneja}.} \bibinfo{year}{2021}\natexlab{}.
\newblock \bibinfo{booktitle}{\emph{2021 Grand Challenge Cheapfakes}}.
\newblock
\urldef\tempurl%
\url{https://github.com/acmmmsys/2021-grandchallenge-cheapfakes}
\showURL{%
\tempurl}


\bibitem[\protect\citeauthoryear{Nguyen, Yamagishi, and Echizen}{Nguyen
  et~al\mbox{.}}{2019}]%
        {Nguyen_2019}
\bibfield{author}{\bibinfo{person}{Huy~H. Nguyen}, \bibinfo{person}{Junichi
  Yamagishi}, {and} \bibinfo{person}{Isao Echizen}.}
  \bibinfo{year}{2019}\natexlab{}.
\newblock \bibinfo{title}{Capsule-forensics: Using Capsule Networks to Detect
  Forged Images and Videos}.
\newblock
\newblock
\showISBNx{9781479981311}
\urldef\tempurl%
\url{https://doi.org/10.1109/icassp.2019.8682602}
\showDOI{\tempurl}


\bibitem[\protect\citeauthoryear{Paris and Donovan}{Paris and Donovan}{2019}]%
        {paris2019}
\bibfield{author}{\bibinfo{person}{Britt Paris} {and} \bibinfo{person}{Joan
  Donovan}.} \bibinfo{year}{2019}\natexlab{}.
\newblock \bibinfo{booktitle}{\emph{{Deepfakes and cheapfakes}: The
  manipulation of audio and visual evidence}}.
\newblock
\urldef\tempurl%
\url{https://datasociety.net/wp-content/uploads/2019/09/DataSociety\_Deepfakes\_Cheap\_Fakes.pdf}
\showURL{%
\tempurl}


\bibitem[\protect\citeauthoryear{Rannard}{Rannard}{2020}]%
        {australia_fire}
\bibfield{author}{\bibinfo{person}{Georgina Rannard}.}
  \bibinfo{year}{2020}\natexlab{}.
\newblock \bibinfo{booktitle}{\emph{Australia fires: Misleading maps and
  pictures go viral}}.
\newblock
\urldef\tempurl%
\url{https://www.bbc.com/news/blogs-trending-51020564}
\showURL{%
Retrieved January 7, 2020 from \tempurl}


\bibitem[\protect\citeauthoryear{Rössler, Cozzolino, Verdoliva, Riess, Thies,
  and Nießner}{Rössler et~al\mbox{.}}{2019}]%
        {ff_dataset}
\bibfield{author}{\bibinfo{person}{Andreas Rössler}, \bibinfo{person}{Davide
  Cozzolino}, \bibinfo{person}{Luisa Verdoliva}, \bibinfo{person}{Christian
  Riess}, \bibinfo{person}{Justus Thies}, {and} \bibinfo{person}{Matthias
  Nießner}.} \bibinfo{year}{2019}\natexlab{}.
\newblock \bibinfo{title}{FaceForensics++: Learning to Detect Manipulated
  Facial Images}.
\newblock
\newblock
\showeprint[arxiv]{1901.08971}~[cs.CV]


\bibitem[\protect\citeauthoryear{Schick}{Schick}{2020}]%
        {mit_tech_report}
\bibfield{author}{\bibinfo{person}{Nina Schick}.}
  \bibinfo{year}{2020}\natexlab{}.
\newblock \bibinfo{booktitle}{\emph{Don’t underestimate the cheapfake}}.
\newblock
\urldef\tempurl%
\url{https://www.technologyreview.com/2020/12/22/1015442/cheapfakes-more-political-damage-2020-election-than-deepfakes/}
\showURL{%
Retrieved Dec 22, 2020 from \tempurl}


\bibitem[\protect\citeauthoryear{Shang, Zhang, Zhang, and Wang}{Shang
  et~al\mbox{.}}{2020}]%
        {Shang2020FauxWardAG}
\bibfield{author}{\bibinfo{person}{Lanyu Shang}, \bibinfo{person}{Yang Zhang},
  \bibinfo{person}{Daniel Zhang}, {and} \bibinfo{person}{D. Wang}.}
  \bibinfo{year}{2020}\natexlab{}.
\newblock \showarticletitle{FauxWard: a graph neural network approach to
  fauxtography detection using social media comments}.
\newblock \bibinfo{journal}{\emph{Social Network Analysis and Mining}}
  \bibinfo{volume}{10} (\bibinfo{year}{2020}), \bibinfo{pages}{1--16}.
\newblock


\bibitem[\protect\citeauthoryear{The White~House}{The White~House}{2014}]%
        {obama_maryland}
\bibfield{author}{\bibinfo{person}{President~Obama The White~House}.}
  \bibinfo{year}{2014}\natexlab{}.
\newblock \bibinfo{booktitle}{\emph{President Obama tours a lab at the Vaccine
  Research Center at the National Institutes of Health}}.
\newblock
\urldef\tempurl%
\url{https://obamawhitehouse.archives.gov/photos-and-video/photo/2014/12/president-obama-tours-lab-vaccine-research-center-national-institutes}
\showURL{%
\tempurl}


\bibitem[\protect\citeauthoryear{{Verdoliva}}{{Verdoliva}}{2020}]%
        {verdoliva2020media}
\bibfield{author}{\bibinfo{person}{L. {Verdoliva}}.}
  \bibinfo{year}{2020}\natexlab{}.
\newblock \showarticletitle{Media Forensics and DeepFakes: An Overview}.
\newblock \bibinfo{journal}{\emph{IEEE Journal of Selected Topics in Signal
  Processing}} \bibinfo{volume}{14}, \bibinfo{number}{5}
  (\bibinfo{year}{2020}), \bibinfo{pages}{910--932}.
\newblock
\urldef\tempurl%
\url{https://doi.org/10.1109/JSTSP.2020.3002101}
\showDOI{\tempurl}


\bibitem[\protect\citeauthoryear{Vorhaus}{Vorhaus}{2020}]%
        {forbes_report}
\bibfield{author}{\bibinfo{person}{Mike Vorhaus}.}
  \bibinfo{year}{2020}\natexlab{}.
\newblock \bibinfo{booktitle}{\emph{People Increasingly Turn To Social Media
  For News}}.
\newblock
\urldef\tempurl%
\url{https://www.forbes.com/sites/mikevorhaus/2020/06/24/people-increasingly-turn-to-social-media-for-news/}
\showURL{%
Retrieved June 24, 2020 from \tempurl}


\bibitem[\protect\citeauthoryear{Wang, Wang, Owens, Zhang, and Efros}{Wang
  et~al\mbox{.}}{2019}]%
        {wang2019detecting}
\bibfield{author}{\bibinfo{person}{Sheng-Yu Wang}, \bibinfo{person}{Oliver
  Wang}, \bibinfo{person}{Andrew Owens}, \bibinfo{person}{Richard Zhang}, {and}
  \bibinfo{person}{Alexei~A Efros}.} \bibinfo{year}{2019}\natexlab{}.
\newblock \showarticletitle{Detecting Photoshopped Faces by Scripting
  Photoshop}. In \bibinfo{booktitle}{\emph{ICCV}}.
\newblock


\bibitem[\protect\citeauthoryear{Wang, Ma, Jin, Yuan, Xun, Jha, Su, and
  Gao}{Wang et~al\mbox{.}}{2018}]%
        {Wang_eann_18}
\bibfield{author}{\bibinfo{person}{Yaqing Wang}, \bibinfo{person}{Fenglong Ma},
  \bibinfo{person}{Zhiwei Jin}, \bibinfo{person}{Ye Yuan},
  \bibinfo{person}{Guangxu Xun}, \bibinfo{person}{Kishlay Jha},
  \bibinfo{person}{Lu Su}, {and} \bibinfo{person}{Jing Gao}.}
  \bibinfo{year}{2018}\natexlab{}.
\newblock \showarticletitle{EANN: Event Adversarial Neural Networks for
  Multi-Modal Fake News Detection}. In \bibinfo{booktitle}{\emph{Proceedings of
  the 24th ACM SIGKDD International Conference on Knowledge Discovery}}
  (London, United Kingdom) \emph{(\bibinfo{series}{KDD '18})}.
  \bibinfo{publisher}{Association for Computing Machinery},
  \bibinfo{address}{New York, NY, USA}, \bibinfo{pages}{849–857}.
\newblock
\showISBNx{9781450355520}
\urldef\tempurl%
\url{https://doi.org/10.1145/3219819.3219903}
\showDOI{\tempurl}


\bibitem[\protect\citeauthoryear{Wikipedia}{Wikipedia}{2021a}]%
        {f1-score}
\bibfield{author}{\bibinfo{person}{Wikipedia}.}
  \bibinfo{year}{2021}\natexlab{a}.
\newblock \bibinfo{booktitle}{\emph{F1-score}}.
\newblock
\urldef\tempurl%
\url{https://en.wikipedia.org/wiki/F-score}
\showURL{%
\tempurl}


\bibitem[\protect\citeauthoryear{Wikipedia}{Wikipedia}{2021b}]%
        {mcc}
\bibfield{author}{\bibinfo{person}{Wikipedia}.}
  \bibinfo{year}{2021}\natexlab{b}.
\newblock \bibinfo{booktitle}{\emph{Matthews correlation coefficient}}.
\newblock
\urldef\tempurl%
\url{https://en.wikipedia.org/wiki/Matthews\_correlation\_coefficient}
\showURL{%
\tempurl}


\bibitem[\protect\citeauthoryear{Yang, Li, and Lyu}{Yang et~al\mbox{.}}{2019}]%
        {deepfake_inconsistent_head_pose}
\bibfield{author}{\bibinfo{person}{Xin Yang}, \bibinfo{person}{Yuezun Li},
  {and} \bibinfo{person}{Siwei Lyu}.} \bibinfo{year}{2019}\natexlab{}.
\newblock \showarticletitle{Exposing Deep Fakes Using Inconsistent Head Poses}.
\newblock \bibinfo{journal}{\emph{ICASSP 2019 - 2019 IEEE International
  Conference on Acoustics, Speech and Signal Processing (ICASSP)}}
  (\bibinfo{date}{May} \bibinfo{year}{2019}).
\newblock
\showISBNx{9781479981311}
\urldef\tempurl%
\url{https://doi.org/10.1109/icassp.2019.8683164}
\showDOI{\tempurl}


\bibitem[\protect\citeauthoryear{Zhang, Shang, Geng, Lai, Li, Zhu, Amin, and
  Wang}{Zhang et~al\mbox{.}}{2018}]%
        {fauxbuster_2018}
\bibfield{author}{\bibinfo{person}{Daniel Zhang}, \bibinfo{person}{Lanyu
  Shang}, \bibinfo{person}{Biao Geng}, \bibinfo{person}{Shuyue Lai},
  \bibinfo{person}{Ke Li}, \bibinfo{person}{Hongmin Zhu},
  \bibinfo{person}{Tanvir Amin}, {and} \bibinfo{person}{Dong Wang}.}
  \bibinfo{year}{2018}\natexlab{}.
\newblock \showarticletitle{FauxBuster: A Content-free Fauxtography Detector
  Using Social Media Comments}. In \bibinfo{booktitle}{\emph{Proceedings of
  IEEE BigData 2018}}.
\newblock


\bibitem[\protect\citeauthoryear{Zlatkova, Nakov, and Koychev}{Zlatkova
  et~al\mbox{.}}{2019}]%
        {zlatkova-etal-2019-fact}
\bibfield{author}{\bibinfo{person}{Dimitrina Zlatkova},
  \bibinfo{person}{Preslav Nakov}, {and} \bibinfo{person}{Ivan Koychev}.}
  \bibinfo{year}{2019}\natexlab{}.
\newblock \showarticletitle{Fact-Checking Meets Fauxtography: Verifying Claims
  About Images}. In \bibinfo{booktitle}{\emph{Proceedings of the 2019
  Conference on Empirical Methods in Natural Language Processing and the 9th
  International Joint Conference on Natural Language Processing
  (EMNLP-IJCNLP)}}. \bibinfo{publisher}{Association for Computational
  Linguistics}, \bibinfo{address}{Hong Kong, China},
  \bibinfo{pages}{2099--2108}.
\newblock
\urldef\tempurl%
\url{https://doi.org/10.18653/v1/D19-1216}
\showDOI{\tempurl}


\end{thebibliography}
